\def\ltsima{$\; \buildrel < \over \sim \;$}
\def\lsim{\lower.5ex\hbox{\ltsima}}
\def\gtsima{$\; \buildrel > \over \sim \;$}
\def\gsim{\lower.5ex\hbox{\gtsima}}
\def\mdot {\dot M}
\def\deg {^\circ}
\def\ergs {~erg$\,$s$^{-1}$}
\def\cms {~cm$\,$s$^{-1}$}
\def\cmdue {~cm$^{-2}$}
\def\cmtre {~cm$^{-3}$}
\def\flux {~erg$\,$s$^{-1}$\,cm$^{-2}$}
\def\cs {~counts\,s$^{-1}$}
\def\gs {~g$\,$s$^{-1}$}
\def\gcm {~g~cm$^{-3}$}
\def\msole{~M_{\odot}}
\def\rsole{~R_{\odot}}
\documentclass{article}
\usepackage{emulateapj,amsfonts,psfig,epsfig}
\begin{document}
\def\ltsima{$\; \buildrel < \over \sim \;$}
\def\lsim{\lower.5ex\hbox{\ltsima}}
\def\loe{\lower.5ex\hbox{\ltsima}}
\def\gtsima{$\; \buildrel > \over \sim \;$}
\def\gsim{\lower.5ex\hbox{\gtsima}}
\def\goe{\lower.5ex\hbox{\gtsima}}
%----------------------------------------------------------------
\def\rref{\par\noindent\hangindent=1.5truecm}
\def\aa #1 #2 {A\&A #1 #2}
\def\aass #1 #2 {A\&AS #1 #2}
\def\araa #1 #2 {ARA\&A #1 #2}
\def\mon #1 #2 {MNRAS #1 #2}
\def\apj #1 #2 {ApJ #1 #2}
\def\apjss #1 #2 {ApJS #1 #2}
\def\apjl #1 #2 {ApJ #1 #2}
\def\astrj #1 #2 {AsJ #1 #2}
\def\nat #1 #2 {Nature #1 #2}
\def\pasj #1 #2 {PASJ #1 #2}
\def\pasp #1 #2 {PASP #1 #2}
\def\msai #1 #2 {Mem. Soc. Astron. Ital. #1 #2}
\def\ass #1 #2 {Ap. Sp. Science #1 #2}
\def\sci #1 #2 {Science #1 #2}
\def\phrevl #1 #2 {Phys. Rev. Lett. #1 #2}
\def\deg {^\circ}
\def\mdot {\dot M}
\def\kms  {\rm \ km \, s^{-1}}
\def\cms  {\rm \ cm \, s^{-1}}
\def\gs   {\rm \ g  \, s^{-1}}
\def\cmtre {\rm \ cm^{-3}}
\def\cmdue {\rm \ cm^{-2}}
\def\gcmdue {\rm \ g \, cm^{-2}}
\def\gcm  {\rm \ g \, cm^{-3}}
\def\rsole {~R_{\odot}}
\def\msole {~M_{\odot}}
\def\fH {{\cal H}}
\def\op {{\cal K}}
\def\nupa{\vfill\eject\noindent}
\def\der#1#2{{d #1 \over d #2}}
\def\inizio{\2acapo\penalty+10000}
\def\fine{\acapo\penalty-10000\blank}
%====================================================================
\received{~~} \accepted{~~} 
\journalid{}{}
\articleid{}{}

\title{The discovery of quiescent X-ray emission from SAX~J1808.4--3658, the 
transient 2.5~ms pulsar}

\author{L. Stella\altaffilmark{1,5}, S. Campana\altaffilmark{2,5}, S. 
Mereghetti\altaffilmark{3}, D. Ricci\altaffilmark{4}, 
G. L. Israel\altaffilmark{1,5}}

\altaffiltext{1}{
Osservatorio Astronomico di Roma, Via Frascati 33,
I--00040 Monteporzio Catone, Roma, Italy,
e-mail: stella@coma.mporzio.astro.it, gianluca@coma.mporzio.astro.it}

\altaffiltext{2}{
Osservatorio Astronomico di Brera, Via Bianchi 46, I--23807
Merate (Lc), Italy, e-mail: campana@merate.mi.astro.it}

\altaffiltext{3}{
Istituto di Fisica Cosmica ``G. Occhialini", CNR, Via Bassini 15,
I--20133 Milano, Italy, e-mail: sandro@ifctr.mi.cnr.it}

\altaffiltext{4}{
BeppoSAX Science Data Center, Via Corcolle 19, I-00131
Roma, Italy, e-mail: riccid@napa.sdc.asi.it}
\altaffiltext{5}{Affiliated to the International Center for Relativistic
Astrophysics}

\begin{abstract}
We report on a 20 ks BeppoSAX observation of the transient 
2.5~ms X-ray pulsar SAX~J1808.4--3658 in its quiescent state. 
A source at a level of $\sim 3\times 10^{-3}$\cs\ was detected with the 
BeppoSAX/MECS at a position consistent with that of SAX~J1808.4--3658. 
The inferred 0.5--10 keV luminosity was $\sim 2-3\times 
10^{32}$\ergs\ (for a distance of 4~kpc), a value similar to that measured 
from other quiescent low mass X-ray transients hosting an old neutron star. 
This result is discussed in the light of the propeller and radio pulsar 
shock emission models for the quiescent emission of these systems.  
\end{abstract}


\section{Introduction}

SAX~J1808.4--3658 was discovered with the BeppoSAX 
Wide Field Cameras, WFCs, in September 1996 when the source 
underwent a $\sim 20$~day outburst (in't Zand et al. 1998). 
The type I X-ray bursts that were detected during the outburst 
(in't Zand et al. 1998) testify that the source is to be classified 
among soft X-ray transients, SXRTs, hosting an old neutron star 
(for a review see Campana et al. 1998a). A distance of 4~kpc was 
inferred from the peak flux of the X-ray bursts, under the assumption 
that this corresponds to Eddington luminosity. For such a distance the 
September 1996 outburst reached a peak luminosity of $\sim
6\times 10^{36}$\ergs. 

SAX~J1808.4--3658 resumed activity in April 1998 at a maximum level of 
$\sim 4\times 10^{36}$\ergs. Pointed observations with the PCA instrument on 
board RossiXTE revealed a $\sim 4$\% modulation at a frequency 
of $\sim 401$~Hz, the first fast coherent signal in the persistent
emission of a neutron star low mass X-ray binary, LMXRB 
(Wijnands \& van der Klis 1998). Pulse
arrival time measurements yielded an orbital period of $\sim 2$~hr, a
projected semimajor axis of $\sim 63$~light-ms and a mass function of
$\sim 3.8\times10^{-5}\,\,M_{\odot}$, consistent with a very low inclination 
system hosting a light companion (Chakrabarty \& Morgan 1998). 
Evidence for a $\sim 2$\% modulation of the X-ray flux at the orbital period 
was also found (Chakrabarty \& Morgan 1998). 
The RossiXTE PCA and HEXTE detected the source up to $\sim 120$~keV;
its spectrum was dominated by a power law-like component  
with photon index of $\sim 2$ (an excess below 10~keV and 
a very gradual high energy cutoff above 35~keV were also present).
The inferred column density was $N_H \sim 4\times 10^{21}$~cm$^{-2}$ 
(Gilfanov et al. 1998; Heindl \& Smith 1998). 

The evolution of the April 1998 outburst of SAX~J1808.4--3658, as monitored 
with the RossiXTE-PCA down to luminosities of $\sim 2\times 10^{34}$\ergs, 
showed a sudden steepening of the flux decay below $\sim 10^{36}$\ergs
(Gilfanov et al. 1998), reminiscent of the one seen from Aql~X-1 
(Campana et al. 1998b). 
During this outburst
the optical counterpart was unambiguously identified with a variable star 
that brightened to $V=16.6$ and displayed a 0.12~mag 
photometric modulation at the 2~hr orbital period 
(Roche et al. 1998; Giles et al. 1999).
The star's position is R.A.(2000): 
18h 08m 27.54s and DEC(2000): --36$\deg$ 58$'$ 44.3$''$. 
Besides the BeppoSAX/WFC and RossiXTE positions, this is also 
consistent with the position of the radio transient source observed with 
ATCA (Gaensler, Stappers \& Getts 1999).

The much celebrated discovery of 2.5~ms coherent pulsations in the persistent 
emission of SAX~J1808.4--3658 (Wijnands \& van der Klis 1998;
Chakrabarty \& Morgan 1998) is the fruition of a two-decade long effort 
aimed at confirming evolutionary scenarios whereby old neutron stars 
in LMXRBs are among the progenitors of 
millisecond radio pulsars. Periods in the ms range
are predicted to result from secular spin-up by accretion torques.
If the neutron stars of LMXRBs 
are endowed with a magnetic field in the $B\sim 10^7-10^9$~G range, 
these are expected to shine as a recycled 
radio pulsar orbiting a low mass companion, once accretion onto them stops.  
The nearly coherent periodicities that have been detected during type I
bursts from seven LMXRBs provided the first convincing evidence that the
neutron stars in these systems are spinning at periods of 2--3 ms
(Strohmayer et al. 1996; for a review see van der Klis 2000). Yet, only 
indirect arguments support the view   
that these neutron stars possess magnetic fields comparable to
those of millisecond pulsars ($\sim 10^8-10^{9}$~G). 

By virtue of the coherent pulsations in its persistent X-ray flux 
SAX~J1808.4--3658 is currently the {\it only} 
fast spinning accreting neutron star in which the presence of a 
magnotosphere has been firmly established. 
The luminosity variations of the source, like those of most other
X-ray transients, are large and allow to sample a variety of physical 
conditions that are not accessible to persistent sources. Therefore 
SAX~J1808.4--3658 can also provide crucial insights in the
different regimes (accretion, propeller and radio pulsar) that are
expected for weakly magnetic, fast spinning neutron stars subject 
to vastly different mass inflow rates. 
This letter presents the results of a BeppoSAX observation of
SAX~J1808.4--3658, the first to detect the source quiescent X-ray flux.  
A short account of this work was presented by Campana (1999). 

\section{BeppoSAX observation and Results}

A pointed observation of SAX~J1808.4--3658
with the Narrow Field Instruments on board BeppoSAX
took place between 1999 March 17, 18:48 UT and  March 19, 10:20 UT.
Due to a much reduced observational efficiency the effective exposure time 
was only 19503~s. X-ray images were obtained with
two Medium Energy Concentrator Spectrometers, MECS, (position sensitive
proportional counters operating in the 1.3--10~keV band; Boella et al.
1997) and a Low Energy Concentrator Spectrometer, LECS, (a thin window
position sensitive proportional counter with extended low energy
response, 0.1--10~keV; Parmar et al. 1997) each in the focal plane of an
X-ray telescope. These in turn consist of a set of 30 nested conical
mirror shells providing an approximation to a Wolter I type geometry
(Conti et al. 1994). Due to UV contamination problems, the LECS was 
operated only at satellite night time, resulting in a shorter still  
exposure time (5495 s). 

By virtue of their smaller point spread function 
and longer exposure time, the two MECS images afforded 
a considerably higher sensitivity than the LECS.
The two MECS images were summed together and searched for point sources by 
using a sliding cell detection algorithm (Giommi et al. 1992).
The 1.3--10~keV MECS image showed a fairly inhomogeneous and high background,  
likely dominated from diffuse (and/or unresolved) emission in the 
galactic centre region 
(note that SAX~J1808.4--3658 is at $l\sim 355\deg$, $b\sim -8\deg$). 
In order to improve the limiting sensitivity to point sources we accumulated 
a MECS image in the 1.3--4.3 keV energy band, where contamination from 
heavily absorbed emission from close to the galactic centre (and beyond) is 
less pronounced. This choice was also motivated by the fact that 
quiescent neutron star low mass X-ray transients are characterised by  
soft X-ray spectra.  

Only one source was detected in the entire image, at a position of 
R.A.(2000): 18h 08m 33s and DEC(2000): --36$\deg$ 57$'$ 48$''$, with 
a 90\% uncertainty radius of $\sim 1.5$~arcmin due to low statistics 
and background inhomogeneities (source A see Fig. 1). 
The position of this source is fully consistent with 
the optical counterpart of SAX~J1808.4--3658. 
The source count rate was 
$\sim (1.9\pm0.5)\times 10^{-3}$\cs in the 1.3--4.3 keV band  
and $\sim (3.1\pm0.7)\times 10^{-3}$\cs in the 1.3--10 keV band. 
The detection is highly significant  
(chance probability of $\sim 3\times 10^{-7}$ and $\sim 8\times 10^{-6}$
in the 1.3--4.3 and 1.3--10~keV images, respectively). 
By contrast we note that in the 1.3--10 keV MECS image 
two additional sources are marginally detected
the positions of which are in any case inconsistent with   
that of SAX~J1808.4--3658 (some $\sim 30$ arcmin away). 
No significant sources were detected in the LECS, nor in the collimated
Phoswich Detector System, PDS, (15--300~keV, Frontera et al. 1997). 

To estimate the probability of finding by chance in our 
image an X-ray source at a position consistent with SAX J1808.4--3658, 
we evaluated the limiting sensitivity of the central region of our 
SAX MECS. This is $\sim 10^{-13}$ erg cm$^{-2}$ s$^{-1}$. Based on the
$\log{N}-\log{S}$ of X-ray sources close to the galactic center derived
with Einstein IPC (0.15--4.5 keV) by Hertz \& Grindlay (1984), we
estimate that the chance probability of detecting an X-ray source within  
a distance of $\sim 1.5 $~arcmin (the BeppoSAX/MECS error radius) 
from the optical position of SAX~J1808.4--3658
is $\sim 0.8$\%. We therefore conclude that the source A 
represents the quiescent counterpart of SAX~J1808.4--3658. 

Photons were extracted from a circular region of radius of $4'$ centered 
on the position of source A. This corresponds to an encircled energy of 90\%. 
The $\sim 200$ photons extracted in this way were grouped according to 
their PHA values so has to have at least 40 photons per channel. 
We determined that the standard BeppoSAX/MECS background
spectra obtained from blank field observations underestimate the 
background in the vicinity of SAX~J1808.4--3658 by a factor of $\sim 2$.  
Therefore we directly estimated the background by using the photons from the 
annular region between 4.7 and 8~arcmin in the 
BeppoSAX/MECS image of the SAX~J1808.4--3658 field. 
Given the small number of source photons 
($\sim 60$ after background subtraction)
the quality of the MECS spectrum 
is very poor. Any simple single component model that we tried produced a 
good fit; examples of this include: a power law with photon index 
$\Gamma\sim 2.5^{+1.1}_{-1.6}$, a black body with temperature 
$k\,T_{\rm bb}\sim 0.8^{+1.2}_{-0.7}$~keV 
or a thermal bremsstrahlung with temperature of 
$k\,T_{\rm br}\sim 2.8^{+\infty}_{-2.2}$ keV 
(all models are for a column density of $N_H=4\times 10^{21}$~cm$^2$;
uncertainties are 90\% confidence for a single parameter).
All models give a 0.5--10 keV unabsorbed flux of $\sim 1-2\times 
10^{-13}$\flux, which converts to a luminosity of 
$\sim 2-3\times 10^{32}$\ergs.
By adopting the ``canonical'' spectrum of quiescent SXRTs, namely  a
soft black body component ($k\,T_{\rm BB}\sim 0.1-0.3$ keV) plus a power
law hard tail with photon index $\sim 2$ (Campana et al. 1998b; Asai
et al. 1998; Campana et al. 2000), the inferred unabsorbed 0.5--10~keV 
luminosity was also $\sim 2-3\times 10^{32}$\ergs.

Despite the small number of photons, we searched for the spin and orbital 
periodicities.  The photon
arrival times were first corrected to the barycentre of the solar system, 
and then to the barycentre of SAX J1808--3658, by using 
a coherence recovery technique and the  
orbital parameters obtained by Chakrabarty \& Morgan (1998).  
We then searched for the coherent signals by folding the data at the 
relevant period. 
We found no evidence for a modulation at the 2.5 ms spin period, nor 
the 2~hr orbital period. The corresponding upper limits to the amplitude
of these modulations were in both cases close to 100\%. 
Moreover we found no evidence for aperiodic variability on timescales 
comparable to the duration of the BeppoSAX observation. 

\section{Discussion}

The BeppoSAX observation reported here led to the discovery of  
quiescent X-ray emission from SAX J1808.4--3658. 
For a distance of 4 kpc, the inferred 0.5--10 keV luminosity 
is $2-3\times 10^{32}$\ergs, well in the range observed in quiescent  
neutron star SXRTs (see e.g. Campana et al. 1998a). 
Such an X-ray luminosity level is too high to result from 
coronal emission from the companion, likely a main sequence star 
of mass $< 0.17$~M$_\odot$ (see e.g. Bildsten \& Rutledge 2000; 
Campana \& Stella 2000). 

In the absence of detailed spectral information, it is not possible to 
ascertain whether the quiescent X-ray emission of SAX J1808.4--3658 
comprises two distinct spectral components as in   
the best studied neutron star SXRTs  (Aql X-1, Campana et al. 1998b and 
Cen X-4, Asai et al. 1998, Campana  et al. 2000). 
These two components have been modelled in terms of a 
power law with photon index of $\sim 2$ plus a soft thermal-like  
spectrum with an equivalent blackbody temperature in the $0.1-0.3$~keV
range; they contribute comparable fractions of the 0.5--10 keV
luminosity. If the quiescent X-ray spectrum of SAX J1808.4--3658 comprised
only the  soft component, this might originate from the 
the release of thermal energy from the neutron star 
interior heated up during the accretion phases by steady burning, 
unless a pion condensate is present in the neutron star core (Campana 
et al. 1998a; Brown, Bildsten \& Rutledge 1998). 
Detailed modelling of thermal emission from the neutron star surface of a
quiescent SXRT envisages a luminosity in the $10^{32}-10^{33}$\ergs\ range 
and a complex X-ray spectrum emerging from the neutron star atmosphere, 
which might help reconciling the small blackbody equivalent 
radii inferred from the observations ($\sim 1-2$~km) with emission 
for the whole neutron star surface  
(Rutledge et al. 1999). Alternative mechanisms have been proposed
for the origin of the soft thermal-like emission from 
quiescent neutron star SXRTs which  
will not be discussed here (see e.g. Campana \& Stella 2000
and references therein). 
 
On the other hand the analogy with other SXRTs, suggests that 
an additional component with a harder spectrum and comparable luminosity
(the power law-like component) 
likely characterises the quiescent X-ray emission of SAX J1808.4--3658.
Two alternative interpretations were proposed for the origin of this
component, both of which involve the regimes expected 
for a fast rotating weakly magnetic neutron star as the mass inflow rate
decreases by decades at the end of an outburst: these are the regime of 
accretion down to the magnetospheric boundary in the propeller regime, 
or shock emission from the interaction 
of the relativistic radio pulsar wind with the matter inflowing from the
companion star (Stella et al. 1994; Zhang et al. 1998; Menou et al. 1999). 
Being the only neutron star low mass transient for which the presence 
of a magnetosphere is firmly established, SAX~J1808.4--3658 provides a 
unique opportunity to test these interpretations. 
Note that the idea that the soft X-ray component (if present) 
originates from the neutron star cooling is not in 
conflict with either of the models above (or at least with some version of 
them). 

The sudden steepening in the X-ray flux decay of SAX~J1808.4--3658 below 
$\sim 10^{36}$\ergs\ (Gilfanov et al. 1998) is suggestive of the onset of
the centrifugal barrier (see also the case of  Aql~X-1, Campana et al.
1998b). By using simple accretion theory to relate the accretion 
luminosity at the onset of the centrifugal barrier to the neutron star
parameters (e.g. Eq.~1 in Stella et al. 1994) a magnetic field 
of $B\sim 10^8$~G is inferred for a spin period of 2.5~ms.
This value of B is within the range inferred by Psaltis \& Chakrabarty (1999)
($10^8-10^9$~G). 
The fact that the 401~Hz pulsations were still observed for some time  
after this steepening might indicate that low level 
accretion occurs from higher latitudes, even in the propeller regime
(Cui et al. 1998).
The expected luminosity jump from the neutron star accretion regime to the 
magnetospheric accretion regime across the centrifugal barrier is limited 
to a factor of $\Delta_c\sim 3$ for a 2.5~ms spinning neutron star (Corbet
1996).  Simple theory predicts also that the ratio of maximum to minimum 
luminosity released by accretion onto the magnetospheric boundary in the
propeller regime is $\Delta_p \simeq 400$, for a 2.5~ms spinning 
neutron star (Campana \& Stella 2000). Note that both $\Delta_c$ and 
$\Delta_p $ depend mainly on the neutron star spin period and 
are independent of the neutron star magnetic field. Therefore, we conclude
that the maximum accretion luminosity ratio from the beginning of the 
onset of the centrifugal barrier to the bottom of the propeller regime, 
is $\Delta_c \Delta_p \sim 1000 $. 
(We adopted a spherical accretion approximation; similar 
results are obtained in the case of a gas-pressure dominated 
disk interacting with the neutron star magnetosphere, see 
Psaltis \& Chakrabarty 1999).
Any additional luminosity resulting, e.g., from the release of rotational
energy of the neutron star in the propeller regime, would make this ratio 
smaller. 
The observed luminosity ratio between the sudden steepening in the outburst
decay and quiescence in SAX J1808.4--3658 is $\sim 3000$,
seemingly too large of a value for the quiescent emission to be powered 
by accretion onto the neutron star magnetosphere. 
This conclusion depends crucially on the identification of the
steepening of the luminosity decay around $\sim 10^{36}$\ergs\ 
with the onset of the centrifugal barrier, a point that is still uncertain  
despite the similarities with the case of Aql~X-1 (Zhang et al. 1998; 
Campana et al. 1998b). 
Therefore the possibility that the propeller regime applies to the
quiescent state of  SAX~J1808.4--3658 cannot be ruled out. 

In the alternative model, the radio pulsar regime applies and a
fraction $\eta$ of the spin down luminosity $L_{sd}$ can be released in the
shock front between the relativistic pulsar wind and matter outflowing from
the companion star, giving rise to an extended power law spectrum  
(Stella et al. 1994; Tavani \& Arons 1997; Campana et al. 1998a). 
In this case the luminosity ratio between the beginning of the onset of 
the centrifugal barrier and the shock emission regime is 
$\Delta_c \Delta_p\Delta_s \sim 2\times 10^2\eta^{-1}$ for a 2.5 ms spinning 
neutron star (here  $\Delta_s$ is the luminosity ratio between the 
minimum accretion luminosity in the propeller regime and the 
radio pulsar shock emission regime). For $\eta \sim 0.1$, a fairly high but 
still realistic value according to current models (see Tavani 1991), the 
above ratio is easily reconciled with the factor of  $\sim 3000$ separating 
the measured luminosities at the steepening of the decay and in quiescence. 
Differently stated, for the $\sim 10^8$~G neutron star magnetic field deduced 
if the action of the centrifugal barrier begins at $\sim 10^{36}$\ergs, 
a shock luminosity of $\eta\, L_{sd} \sim 2\times 10^{33}\ \eta$\ergs\
would be expected, which for $\eta \sim 0.1$ is in agreement with the 
quiescent X-ray luminosity measured from  SAX~J1808.4--3658.
Currently available data therefore appear to favor somewhat the radio pulsar 
shock emission over the propeller interpretation. 

Decisive evidence in favor of the radio pulsar interpretation could 
be obtained through the detection of a pulsed 2.5~ms signal in the 
quiescent X-ray flux of SAX~J1808.4--3658. High throughput 
X-ray observations would be especially useful in searching for this 
pulsed signal, as well as determining the characteristics of the 
quiescent X-ray spectrum. Searches for pulsed radio emission, while somewhat
eased by the independent knowledge of the spin period, 
are nevertheless hampered  by large and presumably variable dispersion measure
and free-free absorption (see e.g. Campana et al. 1998a).

\section{Acknowledgements}
We thank Paolo Giommi for his help with the optimisation 
of the XIMAGE detection algorithm.
This research made use of SAXDAS linearized and cleaned event
files (Rev.2.0) produced at the BeppoSAX SDC. 
Partial support from ASI grants is acknowledged.

\newpage

\begin{figure*}[t]
\centerline{\epsfbox{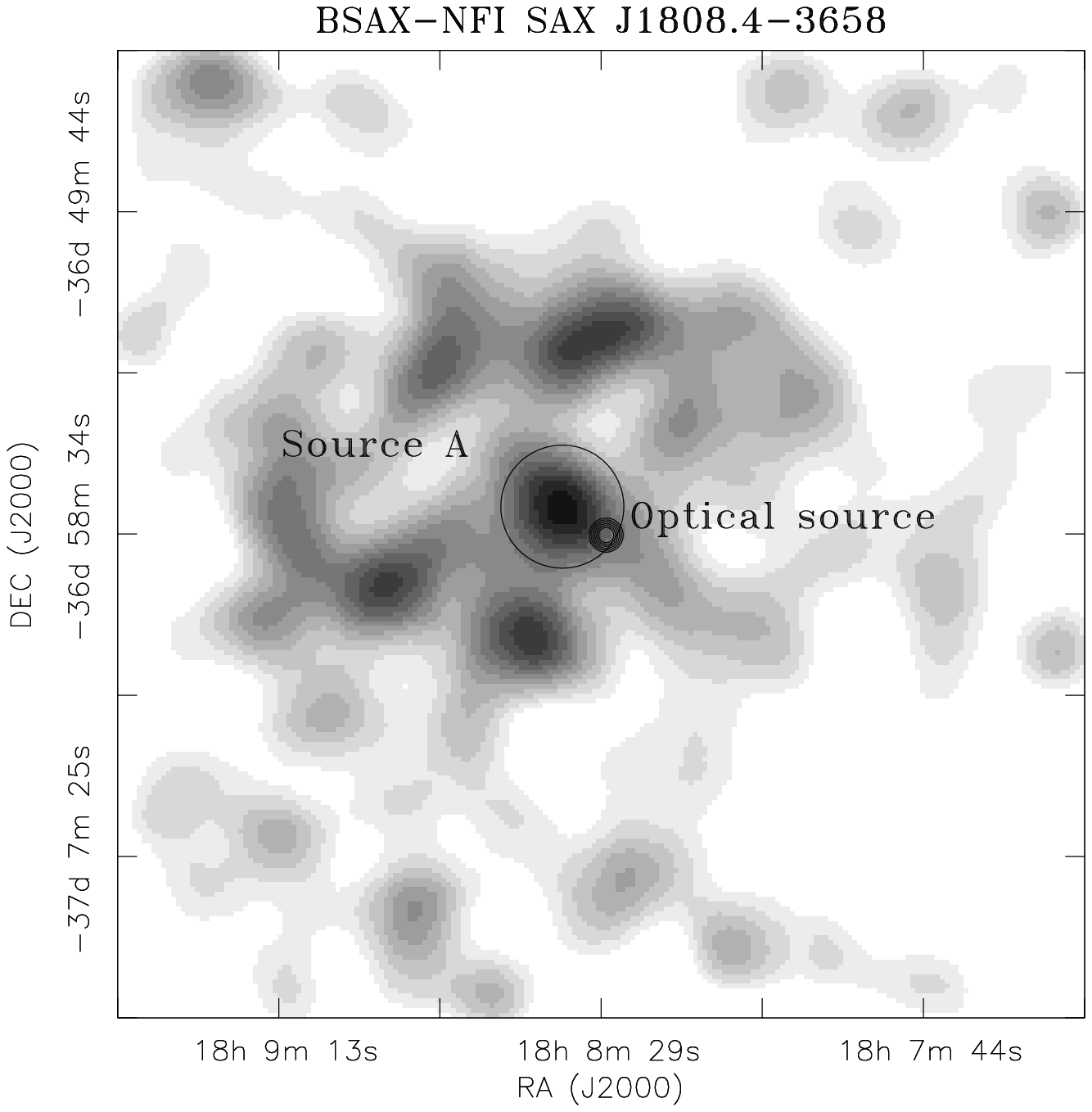}}
\vfill
\begin{center}
{\center Figure 1: Beppo/MECS image of the SAX J1808.4--3658 field in 
the 1.3--4.3~keV energy band. The image was smoothed with a $1.3'$ 
wavelet filter after subtraction of the local background.  
The larger circle represents the SAX J1808.4--3658 error box.
The position of the optical counterpart and radio transient source is 
marked with the smaller circle.}
\end{center}
\end{figure*}

\end{document}